\documentclass[11pt,a4paper]{article}
\usepackage[T1]{fontenc}
\usepackage{lmodern}
\usepackage[a4paper,margin=27mm]{geometry}
\usepackage{amsmath,amssymb,amsthm,mathtools}
\usepackage{microtype}
\usepackage[hidelinks]{hyperref}
\numberwithin{equation}{section}
\theoremstyle{plain}
\newtheorem{theorem}{Theorem}[section]
\newtheorem{proposition}[theorem]{Proposition}
\newtheorem{lemma}[theorem]{Lemma}
\theoremstyle{remark}
\newtheorem{remark}[theorem]{Remark}
\newcommand{\E}{\mathbf E}
\renewcommand{\P}{\mathbf P}
\newcommand{\Q}{\mathbf Q}
\newcommand{\R}{\mathbb R}
\newcommand{\cA}{\mathcal A}
\newcommand{\cL}{\mathcal L}
\newcommand{\cE}{\mathcal E}
\newcommand{\dd}{\,d}

\newcommand{\ArticleTitle}{Optimal Control in Pairs Trading with Negative-Power Utility}
\title{\ArticleTitle}
\author{Aleksei Kozhevnikov\thanks{M. V. Lomonosov Moscow State University and
Vega Institute Foundation, Moscow, Russia.
Email: \href{mailto:aleksei.kozhevnikov@math.msu.ru}{\nolinkurl{aleksei.kozhevnikov@math.msu.ru}}.}}
\date{}
\hypersetup{pdftitle={\ArticleTitle},pdfauthor={Aleksei Kozhevnikov}}

\begin{document}
\maketitle

\begin{abstract}
We solve a finite-horizon consumption and investment problem in a pairs
trading model with an Ornstein--Uhlenbeck spread and utility
$U(x)=-x^{-\gamma}$, $\gamma>0$. Linearizing the
Hamilton--Jacobi--Bellman equation gives explicit formulas for the value
function and the optimal feedback control. A budget inequality and
concavity verify optimality over all admissible strategies with
nonnegative wealth on every finite horizon, without Novikov's condition.
Under the optimal feedback, wealth and consumption admit deterministic
upper bounds when the interest rate $r$ is at most the mean-reversion
rate $\kappa$. If $r>\kappa$, each exceeds any prescribed level with
positive probability at every time after the start.
If $r>\sqrt{2+\gamma}\,\kappa$, there is an
explicit critical horizon at and beyond which expected terminal wealth
is infinite, while optimal expected utility is finite. When the
critical time falls before maturity, expected wealth at that time is
finite exactly when the initial spread is zero. We also show that the
suprema of the negative powers of optimal wealth and consumption are
integrable on every finite horizon exactly for orders $0<m\le 1+\gamma$.
\end{abstract}



\section{Introduction}\label{sec:introduction}

We consider a consumption and investment problem in a pairs trading model
in which the spread follows an Ornstein--Uhlenbeck process
\cite{ElliottVanDerHoekMalcolm2005}. The utility of current consumption
and terminal wealth is $U(x)=-x^{-\gamma}$, with $\gamma>0$.

Albosaily and Pergamenchtchikov~\cite{AlbosailyPergamenchtchikov2024}
studied the consumption and investment problem with an
Ornstein--Uhlenbeck spread and a positive power exponent.
For positive-power utility, Kabanov and
Kozhevnikov~\cite{KabanovKozhevnikov2024} reduced the
Hamilton--Jacobi--Bellman (HJB) equation to a linear equation and solved
the problem on both finite and infinite horizons.
Boguslavsky and Boguslavskaya~\cite{BoguslavskyBoguslavskaya2004}
treated the terminal-wealth problem at zero interest. Boguslavskaya,
Boguslavsky and Muravey~\cite{BoguslavskayaBoguslavskyMuravey2022}
extended that setting to a multivariate Ornstein--Uhlenbeck process.
Mudchanatongsuk, Primbs and Wong~\cite{MudchanatongsukPrimbsWong2008}
solved the HJB equation for terminal utility when the log price
difference follows an Ornstein--Uhlenbeck process. Lipton and L\'opez de
Prado~\cite{LiptonLopezDePrado2020} studied the choice of exit levels for
a position with Ornstein--Uhlenbeck dynamics.

Kim and Omberg~\cite{KimOmberg1996} studied the terminal-wealth problem
when the risk premium follows an Ornstein--Uhlenbeck process.
Wachter~\cite{Wachter2002} obtained an explicit solution to the
consumption problem in a complete market with a mean-reverting risk
premium.

With a negative power exponent, utility tends to minus infinity as
consumption or terminal wealth approaches zero. We use the linearization
from~\cite{KabanovKozhevnikov2024} to construct the feedback control.
We verify its optimality by combining a budget inequality with
concavity, following the martingale approach
\cite{KaratzasLehoczkyShreve1987,CoxHuang1989}. The proof applies to the
full admissible class with nonnegative wealth on every finite horizon,
including horizons on which Novikov's condition fails. The
inverse-moment estimates give the exact range of orders for which the
suprema of negative powers of the optimal processes are integrable on
every finite horizon.

The state-price deflator representation also gives two thresholds for
the interest rate. The explicit deterministic bounds on optimal wealth
and consumption hold when $r\le\kappa$
(Proposition~\ref{prop:boundedness}). If $r>\sqrt{2+\gamma}\,\kappa$,
the expected terminal wealth is infinite at and beyond the critical
horizon~\eqref{eq:H-star}, although the optimal expected utility remains
finite. At the critical time strictly before
maturity, the expected wealth is finite exactly when the initial
spread is zero (Theorem~\ref{thm:finite-means}).

Section~\ref{sec:model} introduces the model and states the main result.
Section~\ref{sec:explicit} solves the HJB equation and derives an
explicit representation of optimal wealth and consumption.
Section~\ref{sec:verification} proves optimality and uniqueness.
Section~\ref{sec:properties} establishes the inverse-moment estimates,
deterministic bounds and criteria for finite expectations.
Appendix~\ref{app:ou-integral} proves the exponential moment formula
for the squared Ornstein--Uhlenbeck integral.

\section{Model and main result}\label{sec:model}

Let $W=(W_v)_{v\ge0}$ be a standard Brownian motion on a complete
probability space $(\Omega,\mathcal F,\P)$. Fix $0\le t<T<\infty$, $x>0$,
and $s\in\R$. Let $B$ be the price of the risk-free asset and $S$ the
price of the risky asset (the spread). Their dynamics are
\[
dB_v=rB_v\dd v,\quad B_t=1,\qquad
dS_v=-\kappa S_v\dd v+\sigma\dd W_v,\quad S_t=s.
\]
Throughout the paper,
\begin{equation}\label{eq:assumptions}
\kappa>0,\qquad \sigma>0,\qquad r\ge0,\qquad
\gamma>0,\qquad \delta\ge0,\qquad \beta>0,
\end{equation}
and no relation between $\kappa$ and $r$ is assumed.
Let $\mathbb F^t=(\mathcal F_v^t)_{v\in[t,T]}$ be the usual augmentation of
the filtration generated by $(W_\theta-W_t)_{t\le\theta\le v}$.
A self-financing portfolio holding $a_v$ units of the risky asset at time
$v$ and consuming at rate $c_v\ge0$ has wealth dynamics
\begin{equation}\label{eq:X}
dX_v=(rX_v-\kappa_1a_vS_v-c_v)\dd v+\sigma a_v\dd W_v,
\qquad X_t=x,\qquad \kappa_1:=\kappa+r.
\end{equation}

A control $(a,c)$ belongs to $\cA_t$ if it is $\mathbb F^t$-predictable,
its paths lie in $L^2([t,T])\times L^1_+([t,T])$ almost surely, and the
corresponding wealth satisfies $X_v\ge0$ for all $v\in[t,T]$ almost surely.

Let $U(y)=-y^{-\gamma}$ for $y>0$ and $U(0)=-\infty$. With the convention
$0^{-\gamma}:=+\infty$, the expected utility and the Bellman function are
\begin{equation}\label{eq:J}
\begin{split}
J(x,s,t;a,c)
&=-\E\left[\int_t^T e^{-\delta(v-t)}c_v^{-\gamma}\dd v
       +\beta e^{-\delta(T-t)}X_T^{-\gamma}\right],\\
J^*(x,s,t)&=\sup_{(a,c)\in\cA_t}J(x,s,t;a,c).
\end{split}
\end{equation}

To simplify the formulas, put
\[
\kappa_\gamma:=\frac{\kappa_1}{1+\gamma},\qquad
\rho:=-\frac{r\gamma+\delta}{1+\gamma}.
\]
We also define
\begin{equation}\label{eq:riccati-data}
D:=\frac{\kappa^2+\gamma r^2}{1+\gamma},\qquad
\lambda_{1,2}:=\frac{\kappa-\gamma r}{1+\gamma}\pm\sqrt D,
\qquad \Delta:=\lambda_1-\lambda_2=2\sqrt D.
\end{equation}
For $t\le v\le\theta\le T$ and $s\in\R$ define
\begin{align}
g^{\theta}(v)&:=\frac{\lambda_1\lambda_2\bigl(1-e^{-\Delta(\theta-v)}\bigr)}
 {\sigma^2\bigl(\lambda_1-\lambda_2e^{-\Delta(\theta-v)}\bigr)},
 \label{eq:g-theta-explicit}\\
f^{\theta}(v)&:=\exp\!\left\{\left(\frac{\lambda_2}{2}+\rho\right)(\theta-v)\right\}
 \sqrt{\frac{\lambda_1-\lambda_2}{\lambda_1-\lambda_2e^{-\Delta(\theta-v)}}},
 \label{eq:f-theta-explicit}\\
\tilde u^{\theta}(s,v)&:=f^{\theta}(v)e^{\frac12g^{\theta}(v)s^2},
 \label{eq:u-theta-explicit}\\
u(s,v)&:=\int_v^T\tilde u^{\theta}(s,v)\dd\theta+\beta^{1/(1+\gamma)}\tilde u^T(s,v).
 \label{eq:u-explicit}
\end{align}

The following theorem gives the Bellman function and the optimal control.

\begin{theorem}\label{thm:main}
Under~\eqref{eq:assumptions}, the Bellman function is
\begin{equation}\label{eq:bellman-final}
J^*(x,s,t)=-x^{-\gamma}u(s,t)^{1+\gamma}.
\end{equation}
Put
\begin{equation}\label{eq:R}
R(s,v):=\frac{u_s(s,v)}{u(s,v)}-\frac{\kappa_\gamma}{\sigma^2}s.
\end{equation}
The equation for $X^*$ with initial condition $X_t^*=x$,
\begin{equation}\label{eq:closed-loop}
dX_v^*=X_v^*\left[\left(r-\kappa_1 S_v R(S_v,v)-\frac{1}{u(S_v,v)}\right)\dd v
                  +\sigma R(S_v,v)\dd W_v\right],
\end{equation}
has a unique strictly positive strong solution. The feedback control
\begin{equation}\label{eq:optimal-feedback}
a_v^*=X_v^*R(S_v,v),\qquad c_v^*=\frac{X_v^*}{u(S_v,v)}
\end{equation}
belongs to $\cA_t$ and is optimal. Optimal controls are unique up to
$dv\otimes d\P$-almost-everywhere equality, and optimal wealth processes
are indistinguishable.

Moreover, for every $0<m\le1+\gamma$, the feedback consumption $c^*$ and
wealth $X^*$ satisfy
\begin{equation}\label{eq:strong-moments}
\E\left[\sup_{t\le v\le T}(X_v^*)^{-m}
       +\sup_{t\le v\le T}(c_v^*)^{-m}\right]<\infty.
\end{equation}
\end{theorem}

Section~\ref{sec:explicit} constructs the candidate optimal control.
Section~\ref{sec:verification} proves optimality and uniqueness, and
Section~\ref{sec:properties} establishes the negative-moment estimate.

\section{Explicit solution of the HJB equation}\label{sec:explicit}

We follow the approach of~\cite{KabanovKozhevnikov2024}. We reduce the HJB
equation to a linear parabolic problem and solve the associated homogeneous
problems by an exponential-quadratic substitution.

\subsection{The HJB equation}

The corresponding Hamilton--Jacobi--Bellman equation on $x>0$ is
\begin{equation}\label{eq:HJB}
\sup_{a\in\R,\,c>0}H(x,s,v,a,c,z)=0,\qquad z(x,s,T)=-\beta x^{-\gamma},
\end{equation}
with the Hamiltonian
\[
H:=z_v+\frac12\sigma^2a^2z_{xx}+\sigma^2az_{xs}+\frac12\sigma^2z_{ss}
+(rx-\kappa_1as-c)z_x-\kappa sz_s-\delta z-c^{-\gamma}.
\]
We shall look for a solution with $z_x>0$ and $z_{xx}<0$. Then the supremum
in $a$ and $c$ is attained at
\begin{equation}\label{eq:feedback-formal}
a^*=\frac{\kappa_1sz_x-\sigma^2z_{xs}}{\sigma^2z_{xx}},
\qquad c^*=\left(\frac{z_x}{\gamma}\right)^{-1/(1+\gamma)}.
\end{equation}

\begin{lemma}\label{lem:linearization}
The function $z(x,s,v)=-x^{-\gamma}u(s,v)^{1+\gamma}$ with a positive
classical solution $u$ of the linear parabolic problem
\begin{equation}\label{eq:u-pde}
u_v+\cL u+1=0,\qquad u(s,T)=\beta^{1/(1+\gamma)},
\end{equation}
where
\begin{equation}\label{eq:generator}
\cL u:=\frac{\sigma^2}{2}u_{ss}
 +(\gamma\kappa_\gamma-\kappa)su_s
 +\left(\rho-\frac{\gamma\kappa_\gamma^2}{2\sigma^2}s^2\right)u,
\end{equation}
is a classical solution of~\eqref{eq:HJB} with $z_x>0$ and $z_{xx}<0$.
The maximum is attained at $a=xR(s,v)$ and $c=x/u(s,v)$.
\end{lemma}

\begin{proof}
We verify the claim by direct substitution. For $u>0$, the signs $z_x>0$
and $z_{xx}<0$ hold, and~\eqref{eq:feedback-formal} gives $a=xR(s,v)$ and
$c=x/u(s,v)$. Substituting these values into the Hamiltonian gives
\[
\sup_{a\in\R,\,c>0}H(x,s,v,a,c,z)
=-(1+\gamma)x^{-\gamma}u^\gamma(u_v+\cL u+1).
\]
The terms in $u_s^2$ arising from $\sigma^2 z_{ss}/2$ and from maximization
over $a$ cancel. At $v=T$, the terminal condition for $u$ gives
$z(x,s,T)=-\beta x^{-\gamma}$.
\end{proof}

\subsection{Solution of the linear equation}

For every $\theta\in[t,T]$, we construct a solution $\tilde u^{\theta}$ of the
homogeneous problem
\begin{equation}\label{eq:u-theta-problem}
\tilde u_v^{\theta}+\cL\tilde u^{\theta}=0,
\qquad
\tilde u^{\theta}(s,\theta)=1.
\end{equation}
Substituting the form~\eqref{eq:u-theta-explicit} into the homogeneous
problem~\eqref{eq:u-theta-problem} gives the equations for the coefficient
of $s^2$ and the constant term,
\begin{align}
\dot g^{\theta}+\sigma^2(g^{\theta})^2+2(\gamma\kappa_\gamma-\kappa)g^{\theta}
-\frac{\gamma\kappa_\gamma^2}{\sigma^2}&=0,
\qquad g^{\theta}(\theta)=0,\label{eq:g-ode}\\
\dot f^{\theta}+\left(\frac{\sigma^2}{2}g^{\theta}+\rho\right)f^{\theta}&=0,
\qquad f^{\theta}(\theta)=1.\label{eq:f-ode}
\end{align}

\begin{proposition}\label{prop:riccati-solution}
One has
\[
D>0,\qquad \lambda_1\lambda_2=-\gamma\kappa_\gamma^2<0,
\qquad \lambda_1>0>\lambda_2.
\]
The functions~\eqref{eq:g-theta-explicit} and~\eqref{eq:f-theta-explicit}
solve~\eqref{eq:g-ode}--\eqref{eq:f-ode}, and
\begin{equation}\label{eq:g-theta-bounds}
\frac{\lambda_2}{\sigma^2}<g^{\theta}(v)\le0,\qquad 0<f^{\theta}(v)\le1,
\qquad t\le v\le\theta\le T.
\end{equation}
\end{proposition}

\begin{proof}
The identities for $D$, $\lambda_1\lambda_2$ and the signs of the roots are
elementary. Putting $q^{\theta}=\sigma^2g^{\theta}$, we rewrite~\eqref{eq:g-ode}
as
\[
\dot q^{\theta}+(q^{\theta}-\lambda_1)(q^{\theta}-\lambda_2)=0,
\qquad q^{\theta}(\theta)=0.
\]
Since $\lambda_1-\lambda_2e^{-\Delta(\theta-v)}>0$, separation of variables
gives~\eqref{eq:g-theta-explicit}, and integration of~\eqref{eq:f-ode}
gives~\eqref{eq:f-theta-explicit}. The bounds for $g^\theta$ follow
from~\eqref{eq:g-theta-explicit} and $\lambda_1>0>\lambda_2$; the bound
$f^\theta\le1$ follows from $\dot f^\theta/f^\theta=-(\sigma^2g^\theta/2+\rho)\ge0$,
since $g^\theta\le0$ and $\rho\le0$, together with $f^\theta(\theta)=1$.
\end{proof}

\begin{theorem}\label{thm:HJB-solution}
The function $u$ defined in~\eqref{eq:u-explicit} is a strictly positive
classical solution of~\eqref{eq:u-pde}, continuous up to $v=T$. More precisely,
with $H:=T-t$,
\begin{equation}\label{eq:u-upper}
0<u(s,v)\le U_T(v):=\int_v^Tf^{\theta}(v)\dd\theta+\beta^{1/(1+\gamma)} f^T(v)
\le H+\beta^{1/(1+\gamma)},
\end{equation}
and
\begin{equation}\label{eq:u-log-derivative}
\frac{u_s(s,v)}{u(s,v)}=s\,\bar g(s,v),
\qquad
\frac{\lambda_2}{\sigma^2}<\bar g(s,v)\le0,
\end{equation}
where $\bar g(s,v)$ is the weighted average of the values $g^\theta(v)$,
$v\le\theta\le T$, with the positive weights in~\eqref{eq:u-explicit}.
Consequently, $z(x,s,v)=-x^{-\gamma}u(s,v)^{1+\gamma}$ is a classical
solution of~\eqref{eq:HJB}, and
\begin{equation}\label{eq:R-growth}
|R(s,v)|\le\frac{|\lambda_2|+\kappa_\gamma}{\sigma^2}|s|.
\end{equation}
\end{theorem}

\begin{proof}
The explicit formulas justify differentiation under the integral
in~\eqref{eq:u-explicit}. Differentiating with respect to $v$ gives the
lower-limit term $-\tilde u^v(s,v)=-1$. By~\eqref{eq:u-theta-problem},
the remaining terms in $u_v+\cL u$ cancel. At $v=T$ the integral vanishes
and $u(s,T)=\beta^{1/(1+\gamma)}$. Positivity and continuity follow from the
formula for $u$, and~\eqref{eq:u-upper} follows from~\eqref{eq:g-theta-bounds}.
Differentiating with respect to $s$ gives~\eqref{eq:u-log-derivative},
which yields~\eqref{eq:R-growth} by~\eqref{eq:R}. Applying
Lemma~\ref{lem:linearization} proves the assertion for $z$.
\end{proof}

\begin{lemma}\label{lem:polynomial-lower}
There exists a constant $c_H>0$, depending only on the parameters
in~\eqref{eq:assumptions} and on $H$, such that
\begin{equation}\label{eq:polynomial-lower}
u(s,v)\ge\frac{c_H}{1+s^2},
\qquad s\in\R,\quad t\le v\le T.
\end{equation}
\end{lemma}

\begin{proof}
The function $-g^{\theta}(v)/(\theta-v)$ depends only on $\theta-v$ and
extends continuously to $\theta=v$, with value $\gamma\kappa_\gamma^2/\sigma^2$
there. Choose $C_H\ge1$ such that
\[
C_H\ge-\rho-\frac{\lambda_2}{2},\qquad
C_H\ge\frac12\sup_{0<\theta-v\le H}\frac{-g^{\theta}(v)}{\theta-v}.
\]
Since $-\dot f^{\theta}/f^{\theta}=\sigma^2g^\theta/2+\rho\ge\lambda_2/2+\rho$,
one has
\[
\tilde u^{\theta}(s,v)\ge e^{-C_H(1+s^2)(\theta-v)},\qquad t\le v\le\theta\le T.
\]
Writing $\tau=T-v$ and $y=1+s^2$ gives
\[
u(s,v)\ge\frac{1-e^{-C_Hy\tau}}{C_Hy}
       +\beta^{1/(1+\gamma)}e^{-C_Hy\tau}\ge\frac{c_H}{y},
\]
where $c_H=\min\{C_H^{-1},\beta^{1/(1+\gamma)}\}$. The second inequality
uses $y\ge1$ and the fact that the two weights $1-e^{-C_Hy\tau}$ and
$e^{-C_Hy\tau}$ sum to one.
\end{proof}

\subsection{Representation of optimal wealth and consumption}

Define the stochastic exponential $Z$ and the state-price deflator $\xi$ by
\begin{equation}\label{eq:deflator}
Z_v:=\cE\left(\frac{\kappa_1}{\sigma}
                  \int_t^\cdot S_\theta\dd W_\theta\right)_v,
\qquad \xi_v:=e^{-r(v-t)}Z_v.
\end{equation}

\begin{lemma}\label{lem:factorization}
Equation~\eqref{eq:closed-loop} has a unique strictly positive strong
solution, and the feedback control~\eqref{eq:optimal-feedback} belongs to
$\cA_t$. With $K=x/u(s,t)$, consumption and wealth have the representations
\begin{equation}\label{eq:deflator-factorization}
c_v^*=K\left(\frac{e^{-\delta(v-t)}}{\xi_v}\right)^{1/(1+\gamma)},
\qquad X_v^*=u(S_v,v)c_v^*.
\end{equation}
Equivalently,
\begin{equation}\label{eq:X-factorized}
\begin{split}
X_v^*&=K u(S_v,v)\exp\!\left\{
C_0(v-t)-C_2\int_t^v S_\theta^2\dd\theta
-\frac{\kappa_\gamma}{2\sigma^2}(S_v^2-s^2)\right\},\\
C_0&:=\frac{r-\delta}{1+\gamma}+\frac{\kappa_\gamma}{2},\qquad
C_2:=\frac{\kappa^2-r^2}{2\sigma^2(1+\gamma)}.
\end{split}
\end{equation}
\end{lemma}

\begin{proof}
The feedback relation~\eqref{eq:optimal-feedback} gives $X_v=u(S_v,v)c_v$.
To construct the candidate, choose consumption to satisfy
\[
e^{-\delta(v-t)}U'(c_v)=U'(K)\xi_v,\qquad c_t=K.
\]
Solving the first equality for $c_v$ and putting $X_v=u(S_v,v)c_v$ gives
the expressions in~\eqref{eq:deflator-factorization}. By It\^o's formula,
\[
\frac{dc_v}{c_v}
=\left(\frac{r-\delta}{1+\gamma}
  +\frac{(2+\gamma)\kappa_\gamma^2}{2\sigma^2}S_v^2\right)\dd v
 -\frac{\kappa_\gamma}{\sigma}S_v\dd W_v.
\]
Let $A_v=(u_s/u)(S_v,v)$. By the product rule and~\eqref{eq:u-pde}, the
diffusion coefficient in $dX_v/X_v$ is
$\sigma A_v-\kappa_\gamma S_v/\sigma=\sigma R(S_v,v)$,
and the drift is
\[
r-\frac{1}{u(S_v,v)}-\kappa_1 S_v A_v
  +\frac{\kappa_1\kappa_\gamma}{\sigma^2}S_v^2
=r-\kappa_1 S_v R(S_v,v)-\frac{1}{u(S_v,v)}.
\]
The term $-\kappa_1 S_v A_v$ is the sum of $-\gamma\kappa_\gamma S_v A_v$
from~\eqref{eq:u-pde} and $-\kappa_\gamma S_v A_v$ from the quadratic
covariation. Thus $X$ solves~\eqref{eq:closed-loop}, with $X_t=x$ by the
choice of $K$. The equation is linear with coefficients continuous along
the paths of $S$, which gives pathwise uniqueness. The constructed $X$
and $c$ are positive and continuous, and $R(S_v,v)$ is continuous.
Consequently the feedback control is admissible.

Finally, It\^o's formula for $S^2$ in~\eqref{eq:deflator} gives
\begin{equation}\label{eq:log-deflator}
\log\xi_v=
\frac{\kappa_1}{2\sigma^2}(S_v^2-s^2)
+\frac{\kappa^2-r^2}{2\sigma^2}\int_t^v S_\theta^2\dd\theta
-\left(r+\frac{\kappa_1}{2}\right)(v-t).
\end{equation}
Substitution in~\eqref{eq:deflator-factorization}
gives~\eqref{eq:X-factorized}.
\end{proof}

\section{Verification}\label{sec:verification}

\subsection{The budget identity}

\begin{proposition}\label{prop:budget}
Every $(a,c)\in\cA_t$ satisfies
\begin{equation}\label{eq:budget-inequality}
\E\left[\xi_TX_T+\int_t^T\xi_vc_v\dd v\right]\le x.
\end{equation}
For the feedback~\eqref{eq:optimal-feedback}, equality holds:
\begin{equation}\label{eq:budget-equality}
\E\left[\xi_TX_T^*+\int_t^T\xi_vc_v^*\dd v\right]=x.
\end{equation}
\end{proposition}

\begin{proof}
The product rule applied to~\eqref{eq:X} and~\eqref{eq:deflator} shows that
\begin{equation}\label{eq:budget-process}
\begin{split}
N_v&:=\xi_vX_v+\int_t^v\xi_\theta c_\theta\dd\theta\\
&=x+\int_t^v\xi_\theta
 \left(\sigma a_\theta+\frac{\kappa_1}{\sigma}S_\theta X_\theta\right)
 \dd W_\theta.
\end{split}
\end{equation}
This representation shows that $N$ is a local martingale.
Since $N\ge0$, it is a supermartingale, which proves~\eqref{eq:budget-inequality}.

For the feedback control,~\eqref{eq:deflator-factorization} gives
\[
N_v^*=K\left[e^{\rho(v-t)}u(S_v,v)Z_v^{\gamma/(1+\gamma)}
       +\int_t^v e^{\rho(\theta-t)}Z_\theta^{\gamma/(1+\gamma)}
          \dd\theta\right].
\]
Since $\rho\le0$ and $u\le H+\beta^{1/(1+\gamma)}$,
\[
\sup_{t\le v\le T}N_v^*
\le K\bigl(2H+\beta^{1/(1+\gamma)}\bigr)
      \left(\sup_{t\le v\le T}Z_v\right)^{\gamma/(1+\gamma)}.
\]
The positive continuous local martingale $Z$ starts at $1$.
Stopping it at its first crossing of a level $y\ge1$, capped at $T$,
gives $\P(\sup_{[t,T]}Z\ge y)\le y^{-1}$. Integration of this tail estimate
yields $\E[(\sup_{[t,T]}Z)^{\gamma/(1+\gamma)}]\le1+\gamma$.
The integrable supremum makes $N^*$ a uniformly integrable martingale
and gives~\eqref{eq:budget-equality}.
\end{proof}

\subsection{Optimality and uniqueness}

We prove the value and uniqueness assertions of Theorem~\ref{thm:main}.
Let $y_0$ be the initial marginal utility of consumption,
\[
y_0:=U'(c_t^*)=\gamma K^{-(1+\gamma)}.
\]
By~\eqref{eq:deflator-factorization} and $u(S_T,T)=\beta^{1/(1+\gamma)}$,
\begin{equation}\label{eq:marginals}
e^{-\delta(v-t)}U'(c_v^*)=y_0\xi_v,
\qquad
\beta e^{-\delta(T-t)}U'(X_T^*)=y_0\xi_T.
\end{equation}
Multiplication by $c_v^*$ and $X_T^*$, respectively, followed
by~\eqref{eq:budget-equality}, gives
\begin{equation}\label{eq:candidate-utility}
\E\left[\int_t^T e^{-\delta(v-t)}(c_v^*)^{-\gamma}\dd v
       +\beta e^{-\delta(T-t)}(X_T^*)^{-\gamma}\right]
=\frac{y_0 x}{\gamma}
=x^{-\gamma}u(s,t)^{1+\gamma}<\infty.
\end{equation}
For $y\ge0$ and $y^*>0$, strict concavity of $U$ gives
\begin{equation}\label{eq:concavity}
U(y)-U(y^*)\le U'(y^*)(y-y^*),
\end{equation}
with equality only when $y=y^*$.

Let $(a,c)\in\cA_t$ and $J(x,s,t;a,c)>-\infty$.
Apply~\eqref{eq:concavity} to consumption and terminal wealth and
use~\eqref{eq:marginals}. The budget terms are integrable by
Proposition~\ref{prop:budget}. Integrating over time and taking
expectations gives
\begin{equation}\label{eq:optimality-comparison}
J(x,s,t;a,c)-J(x,s,t;a^*,c^*)
\le y_0\left(\E\left[\xi_TX_T+
                      \int_t^T\xi_vc_v\dd v\right]-x\right)\le0.
\end{equation}
If $J(x,s,t;a,c)=-\infty$, the inequality is immediate.
From~\eqref{eq:candidate-utility} and~\eqref{eq:optimality-comparison}
we obtain~\eqref{eq:bellman-final}.

For an optimal control, both inequalities in~\eqref{eq:optimality-comparison}
are equalities. Strict concavity of $U$ and Tonelli's theorem then give
$c=c^*$ $dv\otimes d\P$-almost everywhere and $X_T=X_T^*$ almost surely.
The budget equality $\E N_T=N_t$ forces the supermartingale $N$ to have
constant expectation, so $N$ is a martingale.
Since $N_T=N_T^*$, the budget martingales $N$ and $N^*$ are indistinguishable.
By the definition of $N$, equality of consumption and positivity of $\xi$
give indistinguishability of $X$ and $X^*$. By~\eqref{eq:budget-process},
the zero quadratic variation of $N-N^*$ gives
\[
\sigma^2\int_t^T\xi_v^2(a_v-a_v^*)^2\dd v=0\quad\text{a.s.},
\]
so $a=a^*$ $dv\otimes d\P$-almost everywhere.

\section{Moments and boundedness of the optimal wealth}\label{sec:properties}

\subsection{Negative moments}

To prove~\eqref{eq:strong-moments}, it suffices to consider $m=1+\gamma$,
since $y^{-m}\le1+y^{-(1+\gamma)}$ for $y>0$ and $0<m\le1+\gamma$.

In the Ornstein--Uhlenbeck representation
\[
S_v=e^{-\kappa(v-t)}\left(s+
       \sigma\int_t^v e^{\kappa(\theta-t)}\dd W_\theta\right),
\]
the stochastic integral is a martingale with a Gaussian terminal value,
which has all positive moments. Applying Doob's inequality in $L^p$ for
$p>\max\{1,2\ell\}$ gives
\begin{equation}\label{eq:ou-polynomial-moments}
\E\left[\left(1+\sup_{t\le v\le T}S_v^2\right)^\ell\right]<\infty,
\qquad \ell>0.
\end{equation}

By~\eqref{eq:deflator-factorization},
\begin{equation}\label{eq:inverse-consumption}
(c_v^*)^{-(1+\gamma)}
=K^{-(1+\gamma)}e^{(\delta-r)(v-t)}Z_v.
\end{equation}
Together with~\eqref{eq:polynomial-lower}, this gives
\begin{equation}\label{eq:inverse-wealth-bound}
\sup_{t\le v\le T}(X_v^*)^{-(1+\gamma)}
\le C\left(1+\sup_{t\le v\le T}S_v^2\right)^{1+\gamma}
        \sup_{t\le v\le T}Z_v,
\end{equation}
where $C$ is a finite deterministic constant. To bound the right-hand side,
we will show that $\E Z_T^b<\infty$ for some $b>1$. We first record a fact
about stochastic exponentials.

\begin{lemma}\label{lem:linear-exponential}
Let $\hat\kappa\in\R$, let $\vartheta:[t,T]\to\R$ be a bounded
deterministic Borel function, and let $\hat W$ be a Brownian motion.
If $d\hat S_v=-\hat\kappa\hat S_v\dd v+\sigma\dd\hat W_v$ and
$\hat S_t=s$, then
\[
L=\cE\left(\int_t^\cdot\vartheta_\theta\hat S_\theta\dd\hat W_\theta\right)
\]
is a martingale on $[t,T]$.
\end{lemma}

The result follows from the Bene\v{s} criterion
of~\cite[Theorem~4.1 in the Russian original]{KlebanerLiptser2013}.
We give the localization argument for this linear diffusion.

\begin{proof}
For $n>|s|$, let $\tau_n$ be the first time in $[t,T]$ that
$|\hat S|$ reaches $n$, with $\tau_n=T$ if this level is not reached.
The stopped exponential has a bounded integrand, so it defines a
probability measure $\Q_n$ on the terminal sigma-field by
$d\Q_n/d\P=L_{\tau_n}$. Under $\Q_n$, Girsanov's theorem gives the
drift $(-\hat\kappa+\sigma\vartheta_v)\hat S_v$ up to $\tau_n$.
It\^o's formula for $\hat S_{v\wedge\tau_n}^2$ and Gronwall's
inequality give
$\E_{\Q_n}\hat S_{\tau_n}^2\le C$ with $C$ independent of $n$.
Indeed the stopped drift of the square is bounded above by
$C_1(1+\hat S_{v\wedge\tau_n}^2)$ and the stopped stochastic integral
has zero mean. Hence $\Q_n(\tau_n<T)\le C/n^2$. As $L$ is a
nonnegative local martingale with $L_t=1$,
\[
1\ge\E L_T\ge\E[L_T\mathbf 1_{\{\tau_n=T\}}]
=\Q_n(\tau_n=T)\ge 1-C/n^2.
\]
Letting $n\to\infty$ gives $\E L_T=1$. The supermartingale property
then implies that $L$ is a martingale on $[t,T]$.
\end{proof}

\medskip
\noindent\textit{Proof of Theorem~\ref{thm:main}, continued.}
For $b\in\R$, define the measure $\Q^{(b)}$ by its density with respect to
$\P$ on $\mathcal F_T^t$:
\[
\frac{\dd\Q^{(b)}}{\dd\P}
=\cE\left(\frac{b\kappa_1}{\sigma}\int_t^\cdot S_\theta\dd W_\theta\right)_T.
\]
By Lemma~\ref{lem:linear-exponential}, the exponential on the right is a
martingale. Thus the density on $\mathcal F_v^t$ is the same exponential
evaluated at $v$. By Girsanov's theorem, under $\Q^{(b)}$ the drift of
$S$ is $(b\kappa_1-\kappa)S$. Write $\E^{(b)}$ for expectation under
$\Q^{(b)}$ and put $\mu^{(b)}=b(b-1)\kappa_1^2/(2\sigma^2)$. Comparing
$Z_v^b$ with the density on $\mathcal F_v^t$, we obtain
\begin{equation}\label{eq:power-tilt}
\E[Z_v^b F]
=\E^{(b)}\!\left[F\exp\Bigl\{\mu^{(b)}\int_t^v S_\theta^2\dd\theta\Bigr\}\right]
\end{equation}
for every nonnegative $\mathcal F_v^t$-measurable random variable $F$.
The expectations may be infinite.

In~\eqref{eq:power-tilt} take $v=T$, $F=1$ and $b\in(1,2]$. On the fixed
interval $[t,T]$, the means and variances of $S_\theta$ under $\Q^{(b)}$
are bounded uniformly in $\theta\in[t,T]$ and $b\in(1,2]$. By Jensen's
inequality,
\[
\exp\Bigl\{\mu^{(b)}\int_t^T S_\theta^2\dd\theta\Bigr\}
\le\frac1H\int_t^T\exp\{\mu^{(b)}H S_\theta^2\}\dd\theta.
\]
By the exponential moment formula for the square of a Gaussian random
variable, the right-hand side has finite expectation if
\[
2\mu^{(b)}H\sup_{t\le\theta\le T}\operatorname{Var}^{(b)}S_\theta<1,
\]
where $\operatorname{Var}^{(b)}$ denotes variance under $\Q^{(b)}$. Since
$\mu^{(b)}\to0$ as $b\downarrow1$, the condition holds for some $b\in(1,2]$
and $\E Z_T^b<\infty$.

At $b=1$ the density process is $Z$. Thus $Z$ is a martingale and Doob's
inequality gives $\E[(\sup_{t\le v\le T}Z_v)^b]<\infty$ for the chosen
$b>1$. By~\eqref{eq:inverse-consumption}, the supremum of
$(c_v^*)^{-(1+\gamma)}$ is integrable. By~\eqref{eq:ou-polynomial-moments}
and H\"older's inequality with exponents $b/(b-1)$ and $b$, the right-hand
side of~\eqref{eq:inverse-wealth-bound} is integrable. This proves
\eqref{eq:strong-moments} and completes the proof of Theorem~\ref{thm:main}.

\begin{remark}\label{rem:moment-range}
For every $m>1+\gamma$, the left-hand side of~\eqref{eq:strong-moments}
is infinite on all sufficiently long horizons. Put $q_m:=m/(1+\gamma)>1$.
Both $(c_T^*)^{-m}$ and $(X_T^*)^{-m}$ are deterministic positive multiples
of $Z_T^{q_m}$. Apply~\eqref{eq:power-tilt} with $b=q_m$ and $F=1$.
For $b>1$, the coefficient $d_b:=b\kappa_1-\kappa$ is positive. By
Cauchy--Schwarz and the explicit solution for $S$ with drift $d_b S_v$,
\[
\int_t^T S_v^2\dd v\ge\frac1{T-t}\left(\int_t^T S_v\dd v\right)^2,
\qquad
\operatorname{Var}^{(b)}\left(\int_t^T S_v\dd v\right)
=\frac{\sigma^2}{d_b^2}\int_0^{T-t}(e^{d_b u}-1)^2\dd u.
\]
The variance grows exponentially with $T-t$ and hence exceeds
$(T-t)/(2\mu^{(b)})$ for all sufficiently large $T-t$. The exponential
moment of the squared integral with coefficient $\mu^{(b)}/(T-t)$ is
then infinite. It follows from~\eqref{eq:power-tilt} that
$\E Z_T^{q_m}=\infty$.
\end{remark}

\subsection{Boundedness of the optimal consumption and wealth}

\begin{proposition}[Deterministic upper bounds]\label{prop:boundedness}
Assume~\eqref{eq:assumptions}.
\begin{enumerate}
\item[\textup{(i)}] If $\kappa\ge r$, then, simultaneously for $t\le v\le T$,
almost surely,
\begin{align}
0<c_v^*&\le\frac{x}{u(s,t)}
 \exp\!\left\{C_0(v-t)+\frac{\kappa_\gamma s^2}{2\sigma^2}\right\},
 \label{eq:consumption-envelope}\\
0<X_v^*&\le\frac{xU_T(v)}{u(s,t)}
 \exp\!\left\{C_0(v-t)+\frac{\kappa_\gamma s^2}{2\sigma^2}\right\}.
 \label{eq:wealth-envelope}
\end{align}
\item[\textup{(ii)}] If $\kappa<r$, then for every $v\in(t,T]$ and every $M>0$,
\[
\P(X_v^*>M)>0\qquad\text{and}\qquad\P(c_v^*>M)>0.
\]
\end{enumerate}
Thus the optimal wealth and consumption are bounded by deterministic
functions of time if and only if $\kappa\ge r$.
\end{proposition}

\begin{proof}
(i) Here $C_2\ge0$ in~\eqref{eq:X-factorized}. Dropping both nonpositive
random terms in the exponent and using $c_v^*=X_v^*/u(S_v,v)$ gives
\eqref{eq:consumption-envelope}. The inequality $u(S_v,v)\le U_T(v)$
gives~\eqref{eq:wealth-envelope}.

(ii) Assume $\kappa<r$, so that $C_2<0$. Fix $v\in(t,T]$ and put
$I_v=\int_t^v S_\theta\dd\theta$. The pair $(S_v,W_v-W_t)$ has a
nondegenerate Gaussian distribution because the integrands
$\sigma e^{-\kappa(v-\theta)}$ and $1$ are linearly independent on $[t,v]$.
The spread equation gives $\kappa I_v=s-S_v+\sigma(W_v-W_t)$, so the
pair $(S_v,I_v)$ also has a nondegenerate Gaussian distribution. Therefore
the event $\{|S_v|<1,\ |I_v|>L\}$ has positive probability for every $L>0$.
On this event, Cauchy--Schwarz gives
\[
\int_t^v S_\theta^2\dd\theta\ge\frac{I_v^2}{v-t}>\frac{L^2}{v-t},
\]
and $u(S_v,v)\ge\min_{|y|\le1}u(y,v)>0$. As $C_2<0$,
\eqref{eq:X-factorized} and $c_v^*=X_v^*/u(S_v,v)$ imply that both wealth
and consumption exceed $M$ on this event for sufficiently large $L$.
\end{proof}

\subsection{Expected optimal consumption and wealth}

Put
\[
\widetilde\kappa:=\frac{(2+\gamma)\kappa+r}{1+\gamma},\qquad
\zeta:=\sqrt{\frac{r^2-(2+\gamma)\kappa^2}{1+\gamma}}
\quad\text{if }r^2>(2+\gamma)\kappa^2.
\]
The critical horizon is
\begin{equation}\label{eq:H-star}
H_*:=
\begin{cases}
\infty, & r^2\le(2+\gamma)\kappa^2,\\[2pt]
\dfrac{\pi/2+\arctan(\widetilde\kappa/\zeta)}{\zeta}, & r^2>(2+\gamma)\kappa^2.
\end{cases}
\end{equation}

\begin{lemma}[Exponential moment of the squared Ornstein--Uhlenbeck integral]\label{lem:ou-integral}
Let $k>0$, $\mu>0$ and $b\ge0$, let $\hat W$ be a Brownian motion, and let
$\dd\hat S_\theta=-k\hat S_\theta\dd\theta+\sigma\dd\hat W_\theta$,
$\hat S_t=s$. Put $h_b:=\infty$ if $2\sigma^2\mu\le k^2$, and
\[
h_b:=\frac1\nu\Bigl(\frac\pi2+\arctan\frac{k+2\sigma^2 b}{\nu}\Bigr),
\qquad \nu:=\sqrt{2\sigma^2\mu-k^2},
\qquad\text{if }2\sigma^2\mu>k^2.
\]
For $v\in(t,T]$, let
$I_v:=\E\exp\{\mu\int_t^v\hat S_\theta^2\dd\theta-b\hat S_v^2\}\in(0,\infty]$.
\begin{enumerate}
\item[\textup{(i)}] If $v-t<h_b$, then $I_v=\exp\{A(t)+B(t)s^2\}<\infty$,
where $A$ and $B$ solve on $[t,v]$
\[
\dot B-2kB+2\sigma^2B^2+\mu=0,\quad B(v)=-b,
\qquad
\dot A+\sigma^2B=0,\quad A(v)=0.
\]
\item[\textup{(ii)}] If $b=0$ and $v-t\ge h_0$, then $I_v=\infty$.
\item[\textup{(iii)}] If $2\sigma^2\mu>k^2$, $b>0$ and $v-t=h_0$, then
$h_0<h_b$ and
\[
I_v=e^{kh_0/2}\Bigl(\frac{\sqrt{2\mu}}{2\sigma b}\Bigr)^{1/2}
\exp\Bigl\{\Bigl(\frac{k}{\sigma^2}+\frac{\mu}{2\sigma^2 b}\Bigr)s^2\Bigr\}.
\]
\end{enumerate}
\end{lemma}

The proof is given in Appendix~\ref{app:ou-integral}.

\begin{theorem}[Finiteness of the expected optimal consumption and wealth]\label{thm:finite-means}
Assume~\eqref{eq:assumptions}. For each $v\in(t,T]$,
\begin{align*}
\E c_v^*<\infty &\quad\Longleftrightarrow\quad v-t<H_*,\\
\E X_v^*<\infty &\quad\Longleftrightarrow\quad
v-t<H_* \quad\text{or}\quad (v<T,\ v-t=H_*,\ s=0).
\end{align*}
In particular, $\E X_T^*<\infty$ if and only if $T-t<H_*$.
\end{theorem}

\begin{proof}
\textit{Deflator moments.}
For $p>0$, put
\[
\mu_p=\frac{p(1+p)\kappa_1^2}{2\sigma^2},\qquad
k_p=\kappa+p\kappa_1,\qquad
\Delta_p=2\sigma^2\mu_p-k_p^2=p(r^2-\kappa^2)-\kappa^2.
\]
In~\eqref{eq:power-tilt}, take $F=1$ and $b=-p$. Then $\mu^{(-p)}=\mu_p$
and the drift under $\Q^{(-p)}$ is $-k_p S$. By Lemma~\ref{lem:ou-integral}(i),(ii)
with zero terminal coefficient,
\[
\E Z_v^{-p}<\infty\quad\Longleftrightarrow\quad v-t<H(p),\qquad
H(p)=
\begin{cases}
\infty, & \Delta_p\le0,\\[2pt]
\dfrac{\pi/2+\arctan(k_p/\sqrt{\Delta_p})}{\sqrt{\Delta_p}}, & \Delta_p>0.
\end{cases}
\]
For $p=1/(1+\gamma)$, we have $k_p=\widetilde\kappa$ and
$\Delta_p=(r^2-(2+\gamma)\kappa^2)/(1+\gamma)$. Thus
$H(1/(1+\gamma))=H_*$ by~\eqref{eq:H-star}.

\textit{Consumption and terminal wealth.}
By~\eqref{eq:deflator-factorization},
\[
c_v^*=K\left(\frac{e^{(r-\delta)(v-t)}}{Z_v}\right)^{1/(1+\gamma)},
\qquad X_T^*=\beta^{1/(1+\gamma)}c_T^*.
\]
The criterion for $Z_v^{-p}$ with $p=1/(1+\gamma)$ gives the assertions
for consumption and terminal wealth.

\textit{Wealth before and after the critical time.}
Let $v<T$. If $v-t<H_*$, consumption has finite expectation. Since
$X_v^*=u(S_v,v)c_v^*$ and $u\le H+\beta^{1/(1+\gamma)}$, wealth also
has finite expectation. If $v-t>H_*$, then $H_*<\infty$ and the formula
for $H(p)$ shows that it is continuous at $1/(1+\gamma)$. Hence there is
$q\in(0,1)$ such that $v-t>H(q/(1+\gamma))$, and
$\E[(c_v^*)^q]=\infty$. If $\E X_v^*$ were finite,
\eqref{eq:polynomial-lower} and H\"older's inequality would give
\[
\E[(c_v^*)^q]
\le c_H^{-q}(\E X_v^*)^q
\Bigl[\E[(1+S_v^2)^{q/(1-q)}]\Bigr]^{1-q}<\infty
\]
by~\eqref{eq:ou-polynomial-moments}, a contradiction. Thus
$\E X_v^*=\infty$.

\textit{Wealth at the critical time.}
Let $v=t+H_*<T$. Use~\eqref{eq:power-tilt} with $b_*=-1/(1+\gamma)$ and
$F=u(S_v,v)$. Put $\mu_*:=\mu^{(b_*)}$ and $b_\theta:=-g^\theta(v)/2$
for $\theta\in[v,T]$. By~\eqref{eq:u-explicit},
\eqref{eq:deflator-factorization} and Tonelli's theorem,
\[
\E X_v^*=Ke^{(r-\delta)H_*/(1+\gamma)}
\left[\int_v^T f^\theta(v)\Psi(b_\theta)\dd\theta
+\beta^{1/(1+\gamma)}f^T(v)\Psi(b_T)\right],
\]
where
\[
\Psi(b):=\E^{(b_*)}\exp\Bigl\{\mu_*\int_t^v S_\theta^2\dd\theta-bS_v^2\Bigr\},
\qquad b\ge0.
\]
By Lemma~\ref{lem:ou-integral}(iii) with $k=\widetilde\kappa$, $\mu=\mu_*$
and $h_0=H_*$, for $b>0$ we have
\[
\Psi(b)=C_s b^{-1/2}\exp\!\left\{\frac{\mu_* s^2}{2\sigma^2 b}\right\},
\]
where $C_s>0$ is independent of $b$.

For $\theta>v$ we have $b_\theta>0$. Thus the terminal term is finite
and the integrand is continuous on $(v,T]$. By Lemma~\ref{lem:ou-integral}(ii),
$\Psi(0)=\infty$, and the value at $\theta=v$ does not affect the integral.
By~\eqref{eq:g-theta-explicit} and~\eqref{eq:f-theta-explicit},
\[
\lim_{\theta\downarrow v}\frac{b_\theta}{\theta-v}
=\frac{\gamma\kappa_\gamma^2}{2\sigma^2}>0,
\qquad \lim_{\theta\downarrow v}f^\theta(v)=1.
\]
For $s=0$, the integrand is of order $(\theta-v)^{-1/2}$ as
$\theta\downarrow v$, so the integral converges. For $s\ne0$, the integrand
near $v$ is bounded below by
\[
c_1(\theta-v)^{-1/2}\exp\!\left\{\frac{c_2}{\theta-v}\right\},
\qquad c_1,c_2>0,
\]
so the integral diverges. Hence $\E X_v^*<\infty$ if and only if $s=0$.
\end{proof}

By~\eqref{eq:H-star}, $\kappa H_*$ depends only on $r/\kappa$ and $\gamma$.

\begin{remark}\label{rem:wachter}
In our model the Novikov condition in~\cite[equation~(7)]{Wachter2002} is
\[
\E\exp\left\{\frac{\kappa_1^2}{2\sigma^2}\int_t^T S_v^2\dd v\right\}<\infty.
\]
By Lemma~\ref{lem:ou-integral}, it holds on every finite horizon if and only
if $r=0$, since $\kappa_1^2-\kappa^2=r(2\kappa+r)$.

By Theorem~\ref{thm:finite-means}, if $v-t\ge H_*$, then
$\E\xi_v^{-1/(1+\gamma)}=\infty$. Since $y^p\ge y$ for $y\ge1$ and $p\ge1$,
we also have $\E\xi_v^{-p/(1+\gamma)}=\infty$ for every $p\ge1$.
Wachter's sufficient condition~\cite[Appendix~A]{Wachter2002} requires
these moments to be finite for some $p>1$ throughout the horizon. It
fails when $T-t\ge H_*$. Theorem~\ref{thm:main} covers these horizons as well.
\end{remark}

\appendix

\section{Exponential moments of the squared Ornstein--Uhlenbeck integral}\label{app:ou-integral}

\begin{proof}[Proof of Lemma~\ref{lem:ou-integral}]
Put $\tau:=v-\theta$ and $y(\tau):=2\sigma^2B(v-\tau)-k$. The equation
for $B$ becomes
\[
\frac{dy}{d\tau}=y^2+2\sigma^2\mu-k^2,\qquad y(0)=-k-2\sigma^2b .
\]
If $2\sigma^2\mu\le k^2$, the solution increases from $-k-2\sigma^2b$ towards
the equilibrium $-(k^2-2\sigma^2\mu)^{1/2}$, which lies above the initial
value since $\mu>0$; hence it exists for all $\tau\ge0$.

If $2\sigma^2\mu>k^2$, put
$\alpha_b:=\arctan((k+2\sigma^2b)/\nu)$. Separation of variables gives
\begin{equation}\label{eq:riccati-tan}
y(\tau)=\nu\tan(\nu\tau-\alpha_b),\qquad 0\le\tau<h_b.
\end{equation}
The solution tends to $+\infty$ as $\tau\uparrow h_b$.
For $h:=v-t<h_b$ in this case, integration gives
\[
A(t)=\frac12\int_0^h(y(\tau)+k)\dd\tau
=\frac{kh}{2}+\frac12\log\frac{\cos\alpha_b}{\cos(\nu h-\alpha_b)}.
\]
In both cases $B$ and $A(\theta)=\sigma^2\int_\theta^v
B(\ell)\dd\ell$ are bounded on $[t,v]$ when $v-t<h_b$.

(i) Let $v-t<h_b$. It\^o's formula and the equations for $A$ and $B$ give
\[
Y_\theta:=\exp\Bigl\{A(\theta)+B(\theta)\hat S_\theta^2
+\mu\int_t^\theta\hat S_{\ell}^2\dd\ell\Bigr\}
=Y_t\,\cE\Bigl(\int_t^\cdot2\sigma B(\ell)\hat S_{\ell}\dd\hat W_{\ell}\Bigr)_\theta,
\qquad t\le\theta\le v.
\]
Since $B$ is bounded on $[t,v]$, the stochastic exponential is a martingale
by Lemma~\ref{lem:linear-exponential} applied on $[t,v]$.
From $A(v)=0$ and $B(v)=-b$ we obtain $I_v=\E Y_v=Y_t=\exp\{A(t)+B(t)s^2\}$.

(ii) Let $b=0$, $h_0<\infty$ and $v-t\ge h_0$. For $\widetilde v\in(t,t+h_0)$,
denote by $\widetilde A$, $\widetilde B$ the functions of part~(i) with
terminal time $\widetilde v$.
In reversed time for $\widetilde B$ we have $y(0)=-k$ and
$dy/d\tau=y(\tau)^2+\nu^2>0$, so $\widetilde B\ge0$. By part~(i),
\[
I_v\ge I_{\widetilde v}=\exp\{\widetilde A(t)+\widetilde B(t)s^2\}
\ge\exp\{\widetilde A(t)\}.
\]
In the formula for $\widetilde A(t)$ with $b=0$ and $h=\widetilde v-t$, the denominator
$\cos(\nu h-\alpha_0)$ tends to zero as $\widetilde v\uparrow t+h_0$, since
$\nu h_0=\pi/2+\alpha_0$. Thus $\widetilde A(t)\to+\infty$ and $I_v=\infty$.

(iii) Let $2\sigma^2\mu>k^2$, $b>0$ and $v-t=h_0$.
Since $\alpha_b>\alpha_0$, we have $h_0<h_b$, and part~(i) applies.
By~\eqref{eq:riccati-tan}, $y(h_0)=\nu\cot(\alpha_b-\alpha_0)$.
The tangent subtraction formula and $\nu^2+k^2=2\sigma^2\mu$ give
\[
\tan(\alpha_b-\alpha_0)=\frac{b\nu}{\mu+bk},
\qquad
B(t)=\frac{y(h_0)+k}{2\sigma^2}=\frac{k}{\sigma^2}+\frac{\mu}{2\sigma^2b}.
\]
In the formula for $A(t)$ at $h=h_0$,
\[
\frac{\cos\alpha_b}{\cos(\nu h_0-\alpha_b)}
=\frac{\cos\alpha_b}{\sin(\alpha_b-\alpha_0)}
=\frac{1}{(\tan\alpha_b-\tan\alpha_0)\cos\alpha_0}
=\frac{\sqrt{2\mu}}{2\sigma b}.
\]
Substituting the resulting values of $A(t)$ and $B(t)$ into the formula
from~(i) gives~(iii).
\end{proof}

\section*{Acknowledgments}
The author is grateful to the anonymous referee of an earlier version of
this paper for very helpful comments and suggestions.

\section*{Use of AI}
Generative AI tools assisted with proof and reference checks, translation,
and editing.
The author takes full responsibility for the content of the article.

\end{document}